\documentclass{article}

\usepackage{arxiv}

\usepackage[utf8]{inputenc} 
\usepackage[T1]{fontenc}    
\usepackage[hyphens]{url}
\usepackage{hyperref}
\usepackage[hyphenbreaks]{breakurl}
\usepackage{booktabs}       
\usepackage{amsfonts}       
\usepackage{nicefrac}       
\usepackage{microtype}      
\usepackage{graphicx}
\usepackage{natbib}
\usepackage{doi}

\title{Copyright in Generative Deep Learning}

\author{ Giorgio~Franceschelli\\
	Alma Mater Studiorum Università di Bologna\\
	\texttt{giorgio.franceschelli@unibo.it} \\
	\And
	Mirco~Musolesi \\
	University College London\\
	The Alan Turing Institute\\
	Alma Mater Studiorum Università di Bologna\\
	\texttt{m.musolesi@ucl.ac.uk}
}

\date{}

\renewcommand{\headeright}{}
\renewcommand{\undertitle}{}

\hypersetup{
pdftitle={Copyright in Generative Deep Learning},
pdfsubject={q-bio.NC, q-bio.QM},
pdfauthor={Giorgio~Franceschelli, Mirco~Musolesi},
pdfkeywords={Intellectual Property; Deep Learning; Generative Art},
}

\begin{document}
\maketitle

\begin{abstract}
	Machine-generated artworks are now part of the contemporary art scene: they are attracting significant investments and they are presented in exhibitions together with those created by human artists.
    These artworks are mainly based on generative deep learning techniques, which have seen a formidable development and remarkable refinement in the very recent years. Given the inherent characteristics of these techniques, a series of novel legal problems arise. In this article, we consider a set of key questions in the area of generative deep learning for the arts, including the following: is it possible to use copyrighted works as training set for generative models? How do we legally store their copies in order to perform the training process? Who (if someone) will own the copyright on the generated data? We try to answer these questions considering the law in force in both the United States of America and the European Union, and potential future alternatives. We then extend our analysis to code generation, which is an emerging area of generative deep learning. Finally, we also formulate a set of practical guidelines for artists and developers working on deep learning generated art, as well as some policy suggestions for policymakers.
\end{abstract}


\section{Introduction}

In recent years, we have witnessed an exponential growth of interest in the field of generative deep learning (\cite{goodfellow2014, isola, karras2019, yu2016, oord2016wavenet, devlin2018bert, brown2020, ramesh} among others).

Generative deep learning is a subfield of deep learning (\cite{deeplearning}) with a focus on generation of new data. Following the definition provided by \cite{foster}, a generative model describes how a dataset is generated (in terms of a probabilistic model); by sampling from this model, we are able to generate new data.
Nowadays, machine-generated artworks have entered the market (\cite{vernier}); they are fully accessible online\footnote{\url{https://botnik.org/}}; they have the focus of major investments\footnote{\url{https://www.nextrembrandt.com/}}.

Privacy problems in generating faces or personal data are in the spotlight since few years (a technical survey for privacy in visual monitoring systems can be found in \cite{padillalopez}; the use of differential privacy (\cite{dwork}) has opened a new field of research (\cite{xu2019}, \cite{wu2019}, \cite{liu2019}, etc.); but also privacy in clinical data sharing is considered (see, for instance, \cite{beaulieujones})). 
Ethical debates have, fortunately, found a place in the conversation (for an interesting summary of machine learning researches related to fairness, see \cite{chouldechova}), because of biases and discrimination they may cause (as happened with \textit{AI Portrait Ars} (\cite{oleary})), leading to some very remarkable attempts to overcome them, as in \cite{xu2018} or \cite{yu2020inclusive}. 

However, also copyright issues - the focus of this article - arise from the use of deep learning to generate artworks. Generative deep learning techniques are trained on a large amount of data, which may be protected by copyright (especially in case of artworks); moreover, they allow for the generation of new samples similar to the training data, which in turn could claim copyright protection (or could represent plagiarism of other works).
In this context, it is possible to identify at least three problems: the use of protected works, which have to be stored in memory until the end of the training process (even if not for more time, in order to verify and reproduce the experiment); the use of protected works as training set, processed by deep learning techniques through the extraction of information and the creation of a model upon them; the ownership of Intellectual Property rights (if a rightholder would exist) over the generated works.
Although these arguments have already been extensively studied (e.g., \cite{sobel2017} examines use as training set and \cite{deltorn2018} discusses authorship), this paper aims at analyzing all the problems jointly, creating a general overview useful for both the sides of the argument (developers and policymakers); aims at focusing only on generative deep learning, which (as we will see) has its own peculiarities, and not on AI in general (which contains too many different sub-fields that cannot be generalized as a whole); is written by generative deep learning researchers, which may help provide a new and practical perspective to the topic.
For doing so, we first provide the reader with a brief review of what generative deep learning (GDL) means (Section \ref{gen}). We then present an in-depth discussion of the existing legal frameworks in both US and EU in relation to storing protected works for using them as training set (Section \ref{storageandtraining}) and, then, owning the copyright of machine-generated works (Section \ref{ownership}). In addition to legal analysis, we present a policy discussion that leads to practical suggestions for artists and developers working in this new field, as well as to considerations about the adequacy of current laws. Finally, even if the main case study will be about artworks, which are by a long margin the main (potentially copyrighted) outputs generated with deep learning, in the last section we discuss the issues around GDL in relation to source code generation (Section \ref{copilot}).

\section{Generative Deep Learning} \label{gen}

As with other machine learning algorithms, generative deep learning is based on the availability of a training set, which is a set containing examples of the entities that will be our starting point for the generation of new data. The machine learning model itself has to be \textit{probabilistic} in order to generate new data (if it is deterministic, it will generate each time the same entity). 
A generative model is intrinsically different from a discriminative one, which represents the most popular (and classic) technique in machine learning. In discriminative modeling, e.g., classification or logistic regression, we aim at distinguishing among classes after training on labeled data (it is, in a way, a synonymous of supervised learning).  Instead, in generative modeling we use the (unlabeled) dataset to learn to generate new data.

In general, \cite{foster} defines the generative modeling framework as follows: given a dataset of observations $X$, and assuming that $X$ has been generated according to an unknown distribution $p_{data}$, a generative model $p_{model}$ is used to mimic $p_{data}$ in such a way that we can sample from $p_{model}$ to generate observations that appear to have been drawn from $p_{data}$ (and that are suitably different from the observations in $X$).

Generative deep learning is based on the concept of representation learning. The algorithms are used to learn representations of the data that make it easier to extract patterns from them. Differently from classic machine learning techniques, representation learning works directly on unstructured data, not requiring to manually define the features of the input data (necessary to work with structured data). This adds an extra-level of autonomy to the algorithm. Deep learning in itself is based on the composition of multiple non-linear transformations (using multiple stacked layers of processing units) with the goal of yielding more abstract, and ultimately more useful, representations (\cite{bengio2012}). $X$ is typically composed by unstructured data like images or texts.

Several generative models based on deep learning have been proposed in last years (for a more exhaustive list, see \cite{franceschellimusolesi}). However, it is possible to identify a few families of techniques: Variational Auto-Encoders (first proposed by \cite{kingma}, but also used by \cite{gregor2015}); autoregressive models (first proposed by \cite{oord2016}, but also used by \cite{oord2016wavenet, oord2016conditional, parmar2018, devlin2018bert, huang2018music, radford2019, musenet, brown2020}); sequence prediction models (very well explained by \cite{karpathy}, and used by \cite{zhang2014, potash, sturm2016, jacques, lau2018, yi2018, zugarini}); Generative Adversarial Nets (first proposed by \cite{goodfellow2014} and then used by \cite{vernier}; modifications of the first proposal have been done by \cite{yu2016, dong2017, elgammal2017, isola, karras2017, zhang2018, brock2018, karras2019, engel2019}).

We refer the interested reader to the original papers for a detailed description of the technical aspects of generative deep learning given the scope of this article. The first three techniques have in common the fact that the generator is trained directly on the dataset $X$ of examples (with a very small but not important exception for VAE, since the real generator processes only an encoded version of the original work, and not its real version). On the other hand, the generator of a GAN never sees the real examples: it is trained with the so-called adversarial training. A GAN is composed by two networks: a discriminator and a generator (that is, the generative model). The generator never sees real data, while the discriminator is trained on them, with the goal of learning to distinguish between original data and generated ones; at the same time, the generator tries to generate data that fool the discriminator (so, the goal is to generate data that are recognized as real). The learning process proceeds with the two phases alternatively: when the discriminator becomes better in recognizing original data, the generator has to learn more things about their distribution to fool again the discriminator; and when the generator becomes better in generating seemingly real data, the discriminator has to become better in distinguishing them. At the end, the generator should have learned the model correctly, without training on it (\cite{goodfellow2014}). In any case, data are always necessary as in any other machine learning algorithm. We will discuss the implications of this aspect later in this article.

Another very interesting generative deep learning architecture is Google's DeepDream (\cite{mordvintsev}). The designers of DeepDream trained a convolutional neural network for image classification, and then they used the network as a generative model working in reverse: the image is not obtained as an output, but through modifications of an image provided in input, where the modifications are done according to the patterns learned by the network. In particular, the input image is modified in such a way that the ``excitement'' of a selected layer is maximized; this entails the emergence of the patterns that the specific layer has learned to recognize, which might just be simple ornament-like patterns, but also more sophisticated features.


As it is apparent from the overview given above, generative deep learning has very specific characteristics, which have direct implications for copyright.
First of all, generative deep learning techniques are trained on large training sets from which information is extracted. These training sets may contain data, which are protected by copyright, especially if they are artworks. In fact, differently from other learning techniques, they directly learn from the whole work as-it-is. It means researchers need to store potentially protected works in their expressive form, which is what is protected by copyright, and not by the means of features or small chunks if not of their underlined ideas.
Secondly, because of the nature of the generated outputs (especially against the nature of outputs generated by other kinds of artificial intelligence (\cite{colton2008}), which simply follow rules explicitly provided by the programmer). They only depend on the technique used and the knowledge acquired, without any kind of human intervention (except for the coding of the program and the choice of input data), making very difficult to determine who shall be considered as the owner of the copyright.

\section{Storage of Protected Works and Use for Training} \label{storageandtraining}

Our analysis starts considering if the storage, reproduction and therefore the use for training of a protected work by a generative deep learning algorithm violates the copyright, or if it is allowed by US and EU laws.

\cite{sobel2020} identifies four different categories of uses of data performed by machine learning: \begin{enumerate}
  \item uses involving training data not protected by copyright, including works fallen into the public domain (not protected by economic rights anymore);
  \item uses involving copyrighted subject matter released under a permissive license or licensed directly from rightholders;
  \item market-encroaching uses (whose purpose threatens the market of those data);
  \item non-market-encroaching uses (whose purpose is unrelated to copyright's monopoly entitlement).
\end{enumerate}

In the first case, there is no problem in storing and using a work not protected by copyright for this goal. This also applies for works now in the public domain, which happens in EU 70 years after the author's death (or the death of the last of the authors), and in US 95 years after publication date (if created and published before 1978; otherwise, 70 years after the author's death). The same is true also if the work is protected, but it has been acquired digitally through a license agreement that does not expressly prevent a reproduction with this goal (the second case). Otherwise, for protected works on which we have a lawful access but not in digital form or not for a reproduction (third and fourth cases), the question remains open. To address it, the next two subsections consider it under the US law (\ref{us}) and under the EU law (\ref{eu})\footnote{Under Berne Convention, works with a country of origin which is a Union country benefit, in all other Union countries, from the same protection as the latter gives to the works of their own nationals. This means that the protection is governed by the laws of the country where protection is claimed, e.g., a EU research center should be concerned with EU laws.}. Finally, the last subsection considers additional issues related on the \textit{outputs} of generative models, and not on their \textit{inputs} (\ref{additionalissues}).

\subsection{US Law} \label{us}

The US Code establishes that the reproduction of a copyrighted work can be allowed if the use can be considered a fair use of the work (17 U.S.C. § 107 - Limitations of exclusive rights: Fair use). This provision sets the criteria used to determine if the use is fair, i.e., the purpose of the use and its economic character, the nature of the work, the amount and substantiality of the portion used, and the impact of the use over its potential market. 
With these criteria, the law does not state unambiguously what is a fair use and what is not; it provides parameters on which Courts can base their decisions about the fairness of a use. This unpredictability has been criticized, not only because it requires a case-by-case analysis (\cite{netanelbook}) - and eventually to hire a lawyer (\cite{lessig}) - due to its nature of standard more than of rule (\cite{carroll}), but also because the four factors may fail to drive the analysis and may be instead used to support an independent and antecedent conclusion (\cite{nimmer}). However, the fair use doctrine has also some remarkable strengths. It ensures that two competing public interests are balanced: to incentive the creation of new works, and to improve the public's ability to use or access it (see \textit{Sony Corp. v. Universal City Studios, Inc., 464 U.S. 417}). In its form, this doctrine helps exclude uses only where exclusivity promotes social welfare (\cite{lunney}). In addition, even if it seems unpredictable, fair use cases tend to be more coherent than expected and could be organized into clusters, which can help in Courts' decisions (see \cite{samuelson2009}).

Anyway, a deeper analysis of these four criteria in the special case of (generative) deep learning is now necessary.
As regards the amount and substantiality criterion, although in generative deep learning we aim at using the entire work (maybe divided into portions, if it is too long like a novel or a song), we have to consider that machine learning refers to the use of ideas, principles, facts and correlations contained in data given in input; since copyright aims to protect original expressions and not ideas, procedures or methods, data mining techniques do not use copyright works as works \textit{per se} but to access the information stored in them, and so the use is not substantial - and neither they constitute a copyright infringement in theory (\cite{kretschmer}). Also \cite{sag} underlines that a \textit{non-expressive} use such the one discussed above should be considered fair, since it is just about deriving from them meta-level information and not benefiting from their original expression; however, generative techniques could fall under the definition of \textit{expressive} use, since they could use authors' copyrighted expressions (\cite{sobel2017}), learning from their creative and expressive choices (\cite{bonadio}). This would invalidate the insubstantiality theory. In addition, we need to consider that the single protected work is used alongside a large number of other protected works: it is rare that the result resembles one of them substantially, presenting its distinctive features. For this reason, the impact of its potential market is typically very small, because it becomes difficult to connect the generated work with the protected ones used during training (this is true in particular in the case of heterogeneous training sets; of course, if the training set is composed by few works from a single author, this consideration is not more valid). The economic character has to be seen considering that this exception is fair only for purposes like research, and so without a real economic character; however, the previously done distinction between market-encroaching uses and non-market-encroaching uses acquires significance in dividing between an (almost) sure fair use and a dubious case. Finally, when analyzing the purpose of the use (and its fairness), one needs to consider if it is transformative (the most common reason to assess fair use (\cite{asay})): if it adds something new, with a different character, which does not substitute for the original use of the work, the use is more likely to be considered fair\footnote{\url{https://www.copyright.gov/fair-use/more-info.html}}. In particular, the key question to determine a fair use is nowadays if the work is used for a different expressive purpose from that for which the work was created (\cite{netanel}). It is not straightforward to assess this for generative deep learning, but this could eventually be the case of generative deep learning techniques which aim to add a novelty degree in their production (as in \cite{elgammal2017}).
Finally, it is important to highlight that, if the use does not fall in the category of fair use, any sort of reproduction of copyrighted works is not permitted\footnote{\url{https://www.copyright.gov/help/faq/faq-digital.html}}, usually even if it is only on volatile RAM (as judged by many courts, as reported in U.S. Copyright Office, DMCA Section 104 Report 118; however, the Report suggests that in some cases the reproduction of copies on volatile RAM may be considered as a fair use).

As seen, even if without a clear and unambiguous answer, the fair use doctrine certainly offers a support for machine learning researchers (maybe even too much, in comparison with what is allowed to humans (\cite{grimmelmann})). For instance, in the field of Text/Data Mining (TDM), as pointed out by \cite{sag} and confirmed by two famous cases (see \textit{Authors Guild, Inc. v. Google, Inc.} and \textit{Authors Guild, Inc. v. HathiTrust}), the transformative nature of the involved process makes it possible to claim fair use. However, it is not straightforward to extend it to generative deep learning, since the use tends to be more expressive than non-expressive (which makes it more difficult to be considered as fair (\cite{sobel2017})). We can only wait until more clarifications (and Courts' decisions) arrive to unravel it.

\subsection{EU Law} \label{eu}

In the European Union, instead, there are two articles from Directive 2001/29/EC that can or cannot permit the reproduction of protected works.
In particular, Article 2 states there shall be for authors the exclusive right to authorize or prohibit direct or indirect, temporary or permanent reproduction of their work by any means and in any form, in whole or in part; while Article 5(1) provides an exception to this (exclusive) reproduction right (an exception represents a situation in which a right is not reserved - as it usually is). In fact, it states there shall be an exception from the reproduction right in Article 2 for temporary acts of reproduction which are transient or incidental and an integral and essential part of a technological process, and whose sole purpose is to enable a lawful use of a work which has no independent economic significance (therefore excluding market-encroaching uses). This exception is quite similar to the one provided by the US Code, since it requires no independent economic significance (the third case by \cite{sobel2020} is probably excluded; reproductions have not an independent economic significance if they do not generate an additional economic advantage by their use (\cite{triaille}), and it does not seem the case: a neural network, deprived by some of the training data, would probably perform worse) and that the use is lawful (a use is lawful if it is authorized by the rightholder or not restricted by law; we will analyze soon if the use done by generative deep learning is lawful or not), but it adds another constraint: the reproduction must be transient or incidental.
Following \cite{riedo}, this exception could theoretically be applied to Text and Data Mining (TDM) activities (as long as TDM is a lawful use), since during the process data are stored in RAM and then erased when turned off, and so the reproduction is transient. For \cite{schonberger}, Article 5(1) arguably covers copying the works for the training process, deleting them at the end of the process. Despite this, it is very likely that the miners, especially if they are researchers, have to retain the data corpus for verification, aggregation with new data sets and further analysis (\cite{chiou}), deleting it only once their work is completed and published; and therefore storing them until the end of the research is very unlikely to be considered under the exception for temporary reproduction.

To sum up, it is difficult to say that TDM activities are allowed under this exception - or at least there is a degree of uncertainty (\cite{rosati, geiger}). In addition, Article 5(1) requires the use to have no independent economic significance, therefore excluding for-profit research and private companies.
This surely would put the EU's competitive position as both a research and an industrial area in danger (\cite{geiger2015}), and this is one of the main reasons behind the adoption, in 2019, of a Directive ``on copyright and related rights in the Digital Single Market”.

This fundamental Directive tries to answer to all the questions examined above by the means of two Articles. 

Directive's Article 3 states there shall be an exception to allow reproductions and extractions of lawfully accessible protected works for performing text and data mining, if it is made by research organizations and cultural heritage institutions (for the purposes of scientific research and as long as the copies are stored with an appropriate level of security). Notably, the Article states the copies may be retained for the time required for the purposes of scientific research, including for the verification of research results. 
However, we have to remember that the Article only asks for an exception to the \textit{reproduction} right (i.e., the exclusive right to make direct or indirect, temporary or permanent reproduction of the work by any means and in any form) and the \textit{extraction} right (i.e., the exclusive right to permanently or temporarily transfer all or a substantial part of the contents of a database to another medium by any means or in any form). The Article is not about the \textit{making available} right. On the contrary, it is common in scientific research to make source materials available in order to allow others to verify and repeat experiments. Since this is about the making available right and not the reproduction right (\cite{geiger2018}), Directive's Article 3 does not allow to publish protected works used during training.
In principle, this appears to be correct: the researcher has lawful access to the works, while others may not. Making them (or derived versions of them) available means providing others access even if any terms or conditions have not been agreed. However, in practice, this means the verification of research results is not promoted, since it can only be performed by the researchers themselves (\cite{geiger} among others highlighted this in vain before Directive's adoption). In this direction, a good compromise appears to be Article 60d of the German Law on Copyright and Related Rights, which allows the making available of the (normalized and structured) dataset to a ``specifically limited circle of persons for their joint scientific research, as well as to individual third persons'' for quality assurance (\cite{geiger}). Without a statement like this, in case of a research based on not publicly available data, the only (lawful) way to allow the verification of results will be to provide all the necessary information about the data used\footnote{An example is the proposal of Data Cards (\cite{gebru}), whose aim is to address the so-called \textit{documentation debt} (\cite{bandy}), which can also have negative consequences from an ethical perspective (\cite{bolukbasi}). This is impacted by the impossibility of making the dataset available.} and all the pre-processing steps carried out on them or, even better, the related source code.

In addition, Article 4 states there shall also be an exception or limitation to allow reproductions and extractions of lawfully accessible protected works also by other people or institutions, only for the time necessary for the purposes of text and data mining. Crucially, this exception or limitation is applied only if it has not been expressly reserved by their rightholders in an appropriate manner.

To summarize, the Directive includes the use of a (lawfully accessible) protected work for training among the lawful uses: it allows for research organizations to use it for text and data mining. In addition, also other entities can do the same, provided that its rightholder has not expressly reserved this right.

Article 3 is undoubtedly able to foster innovation and scientific research. Even if it only adds an exception for the reproduction right and not for the making available right, making it difficult to reproduce and validate research experiments, in our opinion it is a good compromise between protection and innovation.
Also, Article 4 represents a positive contribution, at least from a theoretical perspective. It is able to encourage innovation also in private environments, avoiding the risk of losing considerable investments (see \cite{hilty} for a discussion of compelling motivations for opening TDM to private entities). In our opinion, leaving the possibility of reserving this exception to rightholders is essential from an ethical perspective (we will talk about it in Section \ref{copilot}). However, many questions arise when trying to operationalise this Article. As highlighted by \cite{rosati2019}, private developers who want to use protected works to train generative models have to follow these three steps: obtain a lawful access to the data; check if rightholders have not reserved the right to make reproductions for TDM purposes; retain any copies made only for as long as is necessary for TDM purposes. The first and the third steps appear to be reasonable; instead, in our opinion, the second presents some issues. How could an EU-based developer know if this right has been reserved for a certain protected work? In Recital 18, the Directive suggests reserving those rights through machine-readable means (e.g., metadata and terms or conditions of a website or a service) in case of publicly available content, and contractual agreements or unilateral declarations for other contents. Even if the list of means that is provided might appear as exhaustive, the issue does not appear to be addressed in a satisfactory manner. As we have seen in Section~\ref{gen}, a generative model is typically trained on a very large dataset. In other words, this might translate into the practical solutions of 1) having online databases that allow to filter (by means of metadata) the available works depending on this reservation, or 2) directly publishing datasets only composed by \textit{reservation-free} works, making sure at the same time they can be integrated with the reserved ones for research purposes. But are these providers forced to do so? Or will this checking activity fall on developers (in a way, dissuading them by training generative models (\cite{chiou}))? Finally, we have also mentioned Google's DeepDream, which only requires one image in input. The same can apply to poems or other texts, which can be fed in input to emerging techniques such as VQ-GAN+CLIP (\cite{vqgan} and \cite{clip}) to obtain \textit{alien dreams} (see \cite{snell}). Here, only one protected work may be used, not extracted by a database provider, but independently acquired. Will the transaction report if this right has been reserved or not? How would it be possible to discover if a work (with an access acquired before 2019) can be used or not? We believe that there is an urgent need of addressing these questions before the TDM exception could be applied in all of its strength.

\subsection{Additional Issues} \label{additionalissues}

We should also remember from Section \ref{gen} that generative deep learning involves the creation of a probabilistic model describing data of interest, from which we obtain new works through sampling. This peculiarity leads to other critical questions. In fact, the model creation and storage, which contains the extracted probabilistic features, may infringe the copyright of works used for training. Its storage cannot be considered as transient or incidental; therefore, it is allowed only if it does not constitute a (partial) reproduction of protected works, since, in that case, there is no copyright relevant activity (\cite{margoni}).

Deep learning models are usually stored as sets of numerical weights; usually, they do not fall in the category of those that are considered as a partial reproduction of a work. However, if the model is built to mimic in output the input (with some non-substantial changes) and it is trained over a protected work, the model would represent that work, at least partially. Moreover, even if the model is not intentionally built to mimic a protected work, it could still end up doing so to an infringing degree: it might reconstruct idiosyncrasies of input data instead of reflecting underlying trends about them (\cite{sobel2017}). If just trained with the goal of learning how to reproduce works, an overfitted generative model may actually be considered as a direct reproduction of the works.

These issues are not only related to the model itself, but also to the output it can generate. Protected input data are commonly used to train models to generate similar output. Then that output may infringe copyright in the pre-existing work or works to which it is similar to (\cite{sobel2017}). In this context, the importance of adversarial training from GANs and searching for diverging from existing works as done by \cite{elgammal2017} may tip the balance towards legality. With respect to other classic techniques, the generative part of a GAN never uses protected data; therefore, it is harder to obtain an output representing an expression of protected data. In addition, new techniques with a novelty objective, which tries to increase the distance between outputs and training data, will tend to be more transformative. This may be crucial, since prior appropriation art cases suggest that, if the result is sufficiently transformative, the use may be protected by fair use or may not represent an infringement of the copyright law (\cite{ligon2019}). However, accidental reproduction of protected works in part might happen, requiring the explicit rightholders' authorization, and not only their not reservation (\cite{sturm2019}). For this reason, in addition to use (new) transformative methods, we suggest to conduct experiments about accidental plagiarism that may be caused by the developed system, as done, for instance, by \cite{hadjeres}.

These considerations about the transformative nature of the result seem fundamental to establish a potential copyright infringement in case of using a protected work as the input image to algorithms like Google's DeepDream. If the result of the modifications of the input is an image that substantially resembles it, it is likely to be considered a reproduction in part, and so this might lead to a copyright infringement. However, as reported by \cite{guadamuz}, this process typically results in producing new images that do not resemble the original ones. This opens the possibility of considering them sufficiently transformative to not be considered as a reproduction in part. In addition, the fact that they are not the result of creative decisions by the programmers leads to the question that we will try to address in the next section.

\section{Copyright of Generated Works} \label{ownership}

The remaining question is who, if anyone, would be the owner of the Intellectual Property rights associated to an artwork produced by a generative model. To answer this question, this section is divided in two parts: a subsection with the current legal analysis (\ref{legalanalysis}); and a subsection with possible future addresses and some policy suggestions (\ref{policysuggestions}).

\subsection{Legal Analysis} \label{legalanalysis}

At first, it is important to make some distinctions. If the generative model is used just as a tool (or the human has an active role in the creative process), the human will be considered as the author; this means that if the human is in charge of the intellectual creation (for instance, by setting all the parameters required to characterize the product, as it is possible with StyleGAN (\cite{karras2019})) or the product can be considered as a co-creation, then the authorship is assigned to that person. In addition, even if the machine has generated the work independently from the human but the latter has selected and evaluated the outputs, rejecting some works and choosing only the best ones following his/her aesthetic tastes, the human can arguably be considered as the author of the work (\cite{glasser}).
As far as works that are fully attributable to a machine are concerned, actually no one would obtain the copyright of machine-generated artworks (\cite{santos2020}). In fact, a fundamental requirement for the application of all the actual laws is that of originality. Even if it is not straightforward to find a precise and applicable definition of originality, in EU it has been commonly considered as satisfied when the work is the reflection of author's personality (\cite{deltorn}), while in US it could be interpreted as a minimum requiring evidence of a human (intellectual) creativity (\cite{gervais2002}). It is questionable to say that computer-generated artworks are the result of the personality of someone - or something - leaving the works unprotected\footnote{This is one of the reasons why WIPO, the World Intellectual Property Organization, has recently started a Conversation on Intellectual Property and Artificial Intelligence to discuss the impact of AI on IP. See for example: \url{https://www.wipo.int/about-ip/en/artificial\_intelligence/conversation.html}}. As a confirmation of this, The Compendium of U.S. Copyright Office Practices establishes that it will not register works produced by a machine or mere mechanical process that operates randomly or automatically without any creative input or intervention from a human author (see Article 313.2), citing, as example, a list of mechanical activities that are the exact opposite with respect to those performed by generative deep learning, and the ones that might be reasonably considered as creative (\cite{palace}). Also Spain, Germany and Australia have formulated a similar criterion, establishing that only works created by humans can be protected by copyright.

On the contrary, the most famous example of a law article for machine-generated artworks is Section 9(3) of the British Copyright, Designs and Patents Act. This section states that in case of a literary, dramatic, musical or artistic work, which is computer-generated, the author shall be taken to be the person by whom the arrangements necessary for the creation of the work are undertaken (same criterion is also considered by Ireland, New Zealand, Hong Kong, South Africa and India). 
This section has been the subject of an intense debate (see \cite{bond}). There is general agreement that for contemporary machine-generated artworks (the ones created by generative deep learning techniques) is difficult, even if not impossible, to find a person who provides necessary arrangements (\cite{smith2017}).
The current lack of protection and the non-general applicability of British criterion are two of the reasons why the European Parliament and Commission have recently been highlighting the need of a specific law for Intellectual Property rights in case of machine-generated works.

In particular, in 2020, the European Parliament adopted a Resolution ``on intellectual property rights for the development of artificial intelligence technologies'', which follows a 2017 Resolution with recommendations to the Commission on Civil Law Rules on Robotics, the first step in trying to regulate this chaotic field. In 2017, the European Parliament asked a solution to protect and, at the same time, foster innovation to the Commission, in order to overcome the problem of non-allocating rights explained above, supporting a horizontal and technologically neutral approach to intellectual property. This approach appears to have an influence on the content of the most recent Resolution. Here, the Parliament highlights the importance of creating a regulation to protect IPRs in the field of AI, in order to protect innovation, guarantee the legal certainty and build the trust needed to encourage investments in these technologies (points 3 and 6). Then, it suggests to not impart legal personality to AI - so, no rights can be assigned to it (point 13). On the contrary, it recommends, if copyright is considered as the correct protection for AI-generated works, to assign the ownership of rights to the person who prepares and publishes a work lawfully, provided that the technology designer has not expressly reserved the right to use the work in that way (from the Explanatory Statement of Report A9-0176/2020, on which the Resolution is based; point 15 specifies that ownership of rights should only be assigned to natural or legal persons that created the work lawfully, but in our opinion the term ``created'' in the context of AI-generated works is misleading and unclear, and the explanation reported above is more useful). In addition, it clarifies that this position is legally correct and harmonized with the existing law; the only requirement is to consider the condition of originality as satisfied not only if the process is creative, but even when the result is creative. This derives from the assumption that AI-generated creation and \textit{traditional} creation still have the aim of expanding cultural heritage in common, even if the creation takes place by means of a different act.
For this reason, if and when this Resolution is embraced by the European Commission too, the owner of the rights will be the person that has prepared and published the work lawfully.

\subsection{Policy Suggestions} \label{policysuggestions}

Even if the current laws do not contemplate machine-generated works for copyright protection, the matter of right attribution has been hugely discussed, not only in terms of law, but also in terms of ethical implications. 

The position of not assigning copyright in machine-generated works may appear to be convenient at first; indeed, it does not ask to change anything.  It might also help preserve the centrality of human authorship in copyright law (see \cite{mezei}) and stress the importance of what an author should be versus what an author should do (see~\cite{craig}). Another, more practical reason is that a work should receive copyright protection only if an \textit{author} exists; but to be considered so, the work must include a meaning or a message he/she wishes to convey, and this cannot happen if no one is able to predict the output of the program (\cite{boyden}), as in deep learning models (\cite{ginsburg}). Finally, placing computer-generated works in the public domain can help preserve the centrality of humans in creative fields, since protection would be guaranteed only to work with an intellectual human contribution (\cite{palace}).

At the same time, there are also strong reasons to not leave them unprotected. First of all, though consistent with the traditional concept of an author as a person, denying protection is inconsistent with the historically flexible interpretation and application of copyright laws as technology has developed. AI products should be evaluated following this flexible interpretation too (\cite{butler1982}). However, the best motivation for the allocation of ownership interests to someone is that the person should be incentivized not for the ideation and creation of the work in itself, but for its public promotion and for making it possible for the computer to create the work (by writing it, training it or instructing it, see \cite{miller1993}). If the law considers a machine-generated work as incapable of being owned because of the lack of a human author, there is limited incentive for creating them and making them public. On the contrary, this might lead to potentially malicious behavior, e.g., the person that used the algorithm to generate the work might be tempted to lie about the way it was created or change it in order to be considered its author (\cite{samuelson1986}).
Finally, the idea that this would mean to incentivize the proliferation of arts and articles of poor quality, penalizing the role of human artists and journalists (\cite{gervais2020}) does not convince us at all. If, as stated, the protection of computer-generated works would translate into a larger number of mediocre works, then for humans it will be easier to produce works of quality higher that those that have been generated by machines and clearly stand out. In addition, even if the current copyright laws were thought to regulate scarcity of products created by humans (\cite{hurt}) and not abundance (of machine-generated products), leaving all of them in the public domain could cause more damage to human authors. The possibility of using them for free may persuade clients to do so, even if human artworks are qualitatively better\footnote{Note that also the art industry feels art lovers would always prefer handmade arts and crafts: \url{https://www.forbes.com/sites/anniebrown/2021/09/06/is-artificial-intelligence-set-to-take-over-the-art-industry/?sh=78b774c33c50}.}.

As we have seen in the previous subsection, European Parliament seems to agree with this line of thought, proposing to allocate rights to who has prepared and published the work lawfully.
The same conclusion (of European Parliament) can be reached in different ways.
Following \cite{franceschelli}, there can be three main individuals involved in the process: the programmer, the person who provides necessary arrangements, and the user. Notice that even if we will consider them separately, in many cases they are the same person. The programmer is just the person who has written the code for the machine; the person who provides necessary arrangements can be, for example, the individual who provides instructions for the desired output, or necessary information about the work the machine has to generate; and the user is the person that, legitimately (because, for instance, the individual is the owner of the machine or has acquired it with a license), ultimately runs the machine and asks it to generate an artwork. In our opinion (but also of the European Parliament and of \cite{yanisky, bohlen, samuelson1986} among others), the rights should be allocated to the user, that can be considered an alter ego of the ``person who prepares and publishes a work lawfully'' (even if also the person who provides the necessary arrangements can be seen under the definition of who prepares the work). Although, in a way, it seems counterintuitive, there are different ways to reach this conclusion. 

\cite{yanisky} and \cite{bohlen} suggest an analogy with the ownership of economic rights in case of software produced by an employee, the so-called work-made-for-hire doctrine: as the employer is entitled to exercise all economic rights in an employee's computer program (if the creation is part of the scope of his/her employment or it is commissioned by the employer), the user is the person who actually causes the creation of the work. Therefore, it is possible to say that the user has \textit{employed} the computer for his/her creative endeavors. In this way, rights can be allocated preventing works from falling into the public domain regardless of the extent of human creative contribution. Also \cite{hristov} and \cite{gurkaynak} suggest to use the analogy with work-made-for-hire doctrine to allocate rights, but with a potentially different conclusion. In fact, they suggest that the equivalent of the employer has to be the programmer or the owner of the generative program, because they are those who really need to be incentivized, rejecting the chance of assigning rights to the end user. However, their definition of end users is quite different from that considered here: here we consider as users the persons who have lawful access to the generative program, and can lawfully use it to generate works. We assume that they are the owners of the program (who, following \cite{wu1997}, should be the copyright owners), possibly because 1) they are also its developers; or 2) they have acquired it from the developers; or 3) following the same doctrine, the developers are their employees. In any case, they are the lawful owners of its economic rights. Alternatively, we assume that they have acquired it via license. In this way there is no need of an extra economic incentive for the developers (they have already been paid or have chosen to release the product freely) or for the owners (they have already been paid for the license or have chosen to release the product freely). On the contrary, the users have paid for using the generative program or they can use it freely because of a particular license. In general, we believe that, even if the generative program has been published and it is accessible by millions of end users, Terms of Service written by the publisher should regulate this kind of problems. We suggest future publishers combine their generators with Terms of Service in order to clearly identify the owner of the generated products (and the associated terms).

Vice versa, \cite{samuelson1986} (but also see \cite{denicola}) states that even if users may not be the ultimate market-makers for machine-generated works, they are in the same position as traditional authors since they take the initial steps that will bring a work into the marketplace (and into its exterior form). Since society has an interest in making these works available to the public, the most effective solution is to give incentives to users to make them available and accessible to others.

Finally, we can reach the same conclusion also working by elimination. The programmer is responsible of the machine creative abilities and, for other kinds of AI (e.g., rule-based systems), it might seem to be enough to establish the originality requirement - and therefore the ownership - in the programmer (\cite{farr}). However, as regards generative deep learning, he/she just creates the potentiality for the creation of the output, but not its actuality (\cite{samuelson1986}). It would be like trying to assign copyright to the teaching artist of a painter, instead of to the painter himself/herself, or, using the analogy proposed by \cite{ralston}, to claiming a knife manufacturer is more responsible for a murder than the person who wielded the knife. The person who provides necessary arrangements can be difficult to identify, and sometimes the generative model may not have such a person associated to it, due to the complexity of deciding which are necessary arrangements and which are just useful arrangements (\cite{franceschelli}). 
A simple example to understand why it is so difficult to use this definition to assign ownership is now presented. Consider the already mentioned Botnik and its creative keyboard\footnote{\url{https://botnik.org/apps/writer/}}. When we open it, it starts with \textit{John Keats} as the source, and it starts suggesting words according to John Keats' texts on which a neural model was previously trained. Then, I can start selecting each time the word in the first position between suggestions, composing a new (and hopefully creative) text. Notice that we could choose the word among different options, and this selection would mean that we would be recognized as the authors (using Botnik just as a tool), be we did not. Now, which shall be considered the necessary arrangements? The only two things we have done as users were to open the website and compulsively click to the first suggested word. Is it enough to consider our actions as necessary arrangements? Of course not. So, necessary arrangements shall be the ones performed by who has loaded John Keats' poems and trained the network; or maybe the ones performed by who has decided that the preset source was John Keats one. But no poems would arise from the creative keyboard without our incredibly simple operations; and it does not seem reasonable to leave, a priori, the ownership of this kind of machine-generated works to someone who was not involved in the materialization of the work - that is, what the law shall protect. No solution seems reasonable, in these cases, in order to assign the rights to who has provided necessary arrangements; of course there can be other cases in which it might seem the right choice, but it would be better to have a rule with the most general applicability, even if \cite{grimmelmann2016} suggests a case-by-case analysis to deal with the heterogeneity of computer-generated works. For this reason, we discard the person who provides necessary arrangements. On the contrary, allocating rights to the user seems to not have any particular flaws; of course, he/she may not have provided any creative contribution, but, as explained above, this does not seem a valid argumentation. For all of these reasons, we think the user should be considered as the owner of (economic) rights.

Finally, an additional consideration about copyright allocation must be done. One of the most explored creative fields by AI researchers is (video)game design. With respect to generative deep learning, it typically concerns the use of generated images (\cite{tilson}), characters (\cite{jin2017}) or soundtracks\footnote{\url{https://cordis.europa.eu/article/id/421438-ai-composers-create-music-for-video-games}} inside games; in these cases, all the conclusions drawn until now are still valid, and no additional considerations are required. However, a growing application is the so-called procedural content generation, where the game scenarios are dynamically generated during game (\cite{liu2021}). Although this task is technically very similar to image generation (with an additional complexity provided by the need of dynamic adaptation and of complexity growth), an additional consideration about copyright allocation is needed. In fact, in procedural content generation, it is not immediate to identify the user (as we have intended him/her). Of course there is the player, who is a \textit{game} user; but the copyright allocation as stated above is about \textit{generative model} user. In this case, the algorithm which generates the game content is not directly used by a person; it is directly used by the game code, and therefore indirectly by the game programmer, who has employed the generative deep learning techniques not to generate content statically, but dynamically. By considering the programmer of the game code as the user of the generative model, the conclusion drawn during this section remains generally applicable in our opinion (and also according to previous literature, e.g. \cite{bridy}).

\section{Code Generation} \label{copilot}

Finally, a quite different generative deep learning use case is about code generation. In particular, GitHub (and OpenAI) Copilot\footnote{\url{https://copilot.github.com/}} has caused a great debate about copyright implications (\cite{guadamuz2021}). Copilot is an AI system able to auto-complete lines of code, but also to generate entire blocks and functions just providing comments or signatures. It is based on an autoregressive model (very close to GPT-3 (\cite{brown2020})) trained on English texts and source codes from publicly available sources as GitHub's public repositories, but not exclusively.

From a legal perspective, GitHub Copilot introduces some additional complexities that are of particular interest for our analysis of copyright in generative learning. With respect to the acquisition and the storage of works used during training, we note that the fact a work is publicly available does not mean it is in the public domain, or it is released under a permissive license -- see points 1) and 2) in Section \ref{storageandtraining}. For instance, the contents of GitHub's public repositories, which are not associated a license are intended to be under copyright's law -- see points 3) and 4)) in Section \ref{storageandtraining}. The GitHub's Terms of Service states that GitHub can process content shared in public repositories as needed to provide the Service, which includes all the applications, software and products provided by GitHub - and therefore also includes Copilot. However, content from external sources is also used. In order to lawfully exploit these sources, their use must fall under the definition of ``fair use'' (or, in EU, must be considered a ``lawful use''), as highlighted above. GitHub itself claims that training machine learning models on publicly available data is considered fair use across the machine learning community; however, it is not straightforward to confirm so. The fair use is determined based on the four criteria examined in \ref{us}. In this case, the use is not for research purpose, and it seems more expressive than non-expressive. In addition, its economic character is not negligible (Copilot is free to use, but companies may use it). The public availability of the works helps satisfying the second criterion for fair use. Then, the work is entirely used during training, but the substantiality of the use is questionable, as discussed before. Finally, the effect upon the potential market depends on the model itself. If it cannot substantially reproduce an existing source code or, if it can, it is able to identify it and refer the user to the original source, also this fourth condition is satisfied and the use could presumably be seen as a fair use.

We also note that publicly available contents have been released under licenses like GNU GPL, with the goal of protecting freedom\footnote{\url{www.gnu.org/philosophy/open-source-misses-the-point.html.en}}. These licenses are chosen in order to avoid any commercial use of the free software, asking to release the derivative work with the same license, in order to foster freedom (and innovation). This is in clear contradiction with the application of the fair use doctrine in case of training a model which could be used for economic purposes. In this way, the question becomes more about ethics than about law: should fair use doctrine be applicable to deep learning independently from the economic character of the application (as it is: fair use is an on/off switch (\cite{ginsburg2014}) and once established the use is fair, nothing prevents the user to do so) or should developers have the opportunity of reserving the right to this kind of utilization?
The EU appears to be aligned to the second option. As discussed in \ref{eu}, Directive's Article 4 considers this use as lawful, unless the rightholder has not expressly reserved this exception. It is not completely clear how to practically reserve it; in our opinion, an interesting possibility will be to augment current licenses and to deal with this right as with the others: as a license specifies if a commercial use is allowed or not, it can also specify if a training use (not only for research purposes) is allowed or not. Of course, this sort of solution could work in US only if they will decide that this use can be reserved, and that the fair use doctrine is not always applicable.

Another important consideration is related to potential copyright infringement. As observed in \ref{additionalissues}, training on protected works could cause a reproduction in parts of the model or, more probably, in the generated output. It has been shown that Copilot quotes existing content very rarely, and it mostly quotes very common code, for which it is almost impossible to detect a substantial reproduction (\cite{copilot}). However, since in principle it may happen, it is praiseworthy that GitHub and OpenAI are building a tracker to detect the rare instances of code that is repeated from the training set. We hope it will help users in identifying the source and author of the code, so that they can directly use it and check the corresponding license, which might require that certain conditions are satisfied.

Finally, the issue related to the ownership of generated code does not introduce any additional complexity. Currently, Copilot works only as a tool, asking the user to test, review and complete the code, therefore the user is lawfully considered as the author - and as the owner. If, in the future, it will not require active supervision anymore and the program will be considered as the author, then it will just become another example of how the work-made-for-hire doctrine perfectly fits generative deep learning. The user, who also provides necessary arrangements (in the form of comments and signatures), is employing Copilot to perform a task within the scope of its employment, therefore even with a higher degree of autonomy, the user should be considered as the owner of related rights.

\section{Conclusion}

In this article, we have explored the most important problems concerning copyright in relation to generative deep learning, trying to understand how the current laws can or cannot permit some common practices. In particular, in Section \ref{storageandtraining}, we have analyzed if and when, under US and EU law, it is possible to store protected works with the goal of training a generative model; the conclusion is that fair use doctrine and 2019 EU's Directive seem to allow this, but with some reservations about the general applicability (for US) and a few practical obstacles (for EU). Then, in Section \ref{ownership}, after having explained that at the moment a generative deep learning output is not protected by copyright, we have also explored future directions in terms of design of legislative frameworks and how these works shall be protected in near future. 

In conclusion, from a practical point of view, as far as researchers are concerned, we suggest 1) to pay attention at the information that is stored (and for how long) during the whole process (and, in case of private use, to carefully check terms or other conditions which could prevent the TDM exception); 2) to try to diverge from the dataset used during training, in order to avoid the risk of a reproduction in part, but also in order to strengthen a transformative use claim; 3) to clarify their position through Terms of Service if other users would be able to use their generative model; and 4) of course, to keep update about the evolution of the legislative frameworks at national and international level.

\bibliographystyle{apalike}
\bibliography{main}  

\end{document}